\documentclass[11pt]{article}
\usepackage[pdftex]{graphicx}     
\usepackage{amsfonts}
\usepackage{amssymb,amsmath, color}

\newcommand{\bm}{\bf}

\newcommand{\rz}{\mathbb R}
\newcommand{\s}{\mathbb S}
\newcommand{\hz}{\mathbb H}

\def\I{\ifmmode{1 \hskip -5pt \mbox{I}}
    \else{\hbox{$1 \hskip -5pt \mbox{I}$}}\fi}      
\def\g{\ifmmode{g \hskip -7pt g}
    \else{\hbox{$g \hskip -7pt g$}}\fi}         

\newtheorem{prop}{Proposition}
\newtheorem{rem}{Remark}
\newtheorem{definition}{Definition}

\input{epsf}

\begin{document}

\title
{
The Generalized Spherical Radon Transform and Its Application in Texture Analysis
}
\author{Swanhild Bernstein, Ralf Hielscher, Helmut Schaeben}

\begin{center}
\LARGE \bf
The Generalized Spherical Radon Transform and Its Application in Texture Analysis
\end{center}
\bigskip
\vspace{25ex}
\begin{center}
\Large S. Bernstein, R. Hielscher, H. Schaeben\footnote{\large Freiberg University of Mining and Technology, D -- 09596 Freiberg, Germany}
\end{center}

\newpage

\begin{center}
\LARGE\bf
The Generalized Spherical Radon Transform and Its Application in Texture Analysis
\end{center}
\bigskip
\vspace{12.5ex} \large
{\bf MOS subject classification:} 15A66, 51M15, 58D15, 44A12
\vspace{12.5ex}
\begin{abstract}
The generalized spherical Radon transform associates the mean values over spherical tori to a function $f$ defined on $\s^3 \subset \hz$, where the elements of $\s^3$ are considered as quaternions representing rotations. It is introduced into the analysis of crystallographic preferred orientation and identified with the probability density function corresponding to the angle distribution function $W$. Eventually, this communication suggests a new approach to recover an approximation of $f$ from data sampling $W$. At the same time it provides additional clarification of a recently suggested method applying reproducing kernels and radial basis functions by instructive insight in its involved geometry.
The focus is on the correspondence of geometrical and group features but not on the mapping of functions and their spaces.
\end{abstract}

\section{Motivation and Introduction}
In texture analysis, i.e. the analysis of preferred crystallographic
orientation, the orientation probability density function $f$
representing the probability law of random orientations of crystal
grains by volume is a major issue. In x--ray or neutron diffraction
experiments spherical intensity distributions are measured which can
be interpreted in terms of spherical probability distributions of
distinghuished crystallographic axes. In texture analysis, they are
referred to as pole probability density functions. Generally, if $f$
is the orientation probability density function of a random rotation
represented by its random quaternion variable $Q$, then the
probability density function of the random direction $Q{\bm h}
Q^{\ast}$, where ${\bm h} \in \s^2 \subset \rz^3$ is a fixed
crystallographic direction, is provided by the $1$--dimensional
spherical Radon transform, integrating $f$ over all $1$--dimensional
great circles $C \subset \s^3$ representing all rotations mapping
${\bm h}$ onto a direction ${\bm r} \in \s^2$.

Extracting pole densities from raw intensity data collected in
diffraction experiments requires several steps of processing, in
particular defining the ``background intensity'' and correspondingly
normalizing the intensity data. Any step in this process is largely
subject to primary experimental errors. The influence of
experimental errors can easily be reduced by considering mean
intensities resulting from an additional rotation of the specimen
during the measurements (Nikolayev, 2004). These mean values
correspond to integration of spherical pole density functions over
small circles. Thus, they correspond to mean values of $f$ over
$2$--dimensional spherical tori $T \subset \s^3$ with core $C$
(Meister and Schaeben, 2004) which are called generalized spherical
Radon transform (Helgason, 1994; 1999). Here, it is identified with
the angle density function $W$ (cf. Bunge, 1969, 1982).
Independently of the new experimental possibilities, they have
proven to be instrumental for the inverse spherical Radon transform
(Muller et al., 1981;\ Helgason, 1994; 1999).

\section{Geometry of Rotations}
One of the most beautiful features of quaternions is the role they play in the representation of the rotations of the low dimensional spaces
$\mathbb{R}^3$ and $\mathbb{R}^4.$ The description here is based on the book by Delanghe, Sommen and Sou{$\check{{\rm c}}$}ek (1992).

\subsection{Rotations and Quaternions}
The skew-field of quaternions  $\mathbb{H}$ is generated by the elements $e_i,\ i=1,\,2,\,3,$ which fulfil the relations\\
\begin{description}\setlength{\itemsep}{0ex}
    \item[(i)] $e_i^2=-1,\ i=1,\,2,\,3; $
    \item[(ii)] $ e_1e_2=e_3,\ e_2e_3=e_1,\ e_3e_1=e_2;$
    \item[(iii)] $e_ie_j+e_je_i=0,\ i,j=1,\,2,\,3;\ i\not=j.$
\end{description}
The unit element of $\mathbb{H}$ should be denoted by $e_0.$ Then an arbitrary quaternion $q\in\mathbb{H}$ can be represented as
$$ q = q_0+ \mathbf{q}=q_0e_0+\sum_{i=1}^3q_ie_i = {\rm Sc}\,q+{\rm Vec}\,q,    $$
where $q_0={\rm Sc}\,q$ is the scalar part and $\mathbf{q}={\rm Vec}\,q$ is the vector part of $q.$ If ${\rm Sc}q=0,$ then $q$ is called a pure quaternion, the subset of all pure quaternions is denoted by ${\rm Vec}\,\mathbb{H}.$ The subset of all quaternions with vanishing vector part may be denoted by ${\rm Sc}\,\mathbb{H}.$ In this way
$\mathbb{R}^3$ and $\mathbb{R}$ are embedded in $\mathbb{H}$ and $\mathbb{R}^4=\mathbb{R}^3\oplus\mathbb{R}$ may be identified with $\mathbb{H}.$\\[2ex]
The quaternion
$$q^{*}=q_0-\mathbf{q}=q_0e_0-\sum_{i=1}^3q_ie_i = {\rm Sc}\,q-{\rm Vec}\,q $$
is called the conjugate of $q.$ The conjugation is an anti-involution, i.e. $(pq)^{*}=q^{*}p^{*},$  and
$$qq^{*}=q^{*}q=||q||^2=\sum_{j=0}^3q_j^2, $$
where $||q||$ is the norm of the quaternion $q$ which coincides with the norm of $q$ in the Euclidean space $\mathbb{R}^4.$ We denote the set of all unit quaternions, i.e. all quaternions with norm 1, by $\s^3$.\\[2ex]
A linear transform $T$ in $\mathbb{R}^3$ or $\mathbb{R}^4$ is called a \emph{rotation} if and only if it leaves the scalar product in the Euclidean space \emph{and} the orientation invariant. Of course we have that $T$ is invertible with $\det T=1.$ The set of all rotations (in the previous sense) in $\mathbb{R}^3$ and
$\mathbb{R}^4$ is given by the special orthogonal group $\mathbf{SO}(3)$ and $\mathbf{SO}(4)$ respectively.

\begin{prop}\hspace{-5pt}{\bf .}
Let $q\in \s^3$ and define the linear transform $T_q:\,\mathbb{R}^3\to\mathbb{R}^3$ by
$$T_q(x)=qxq^{*}.$$
Then $T\in \mathbf{SO}(3)$ if and only if $T=T_q$ for some $q\in \s^3.$
\end{prop}
Thus any rotation in $\mathbb{R}^3$ can be represented by a unit quaternion $q$ via $qxq^{*}.$ As it turns out, quaternions can also be used to represent rotations in $\mathbb{R}^4,$ i.e. the group $\mathbf{SO}(4).$
\begin{prop}\hspace{-5pt}{\bf .}
Let $q,\,p\in\s^3$ be unit quaternions and define the maps
$$ \Phi_q,\,\Psi_p:\,\mathbb{R}^4\to \mathbb{R}^4 \quad \mbox{by} \quad \Phi_q(x)=qx \quad \mbox{and} \quad \Psi_p(x)=xp.$$
Then $T\in\mathbf{SO}(4)$ if and only if there exist $q,\,p\in\s^3$ such that $T=\Theta_{q,p^{*}},$ where
$$ \Theta_{q,\,p^{*}} = \Phi_q \circ \Psi_{p^{*}} = \Psi_{p^{*}} \circ \Phi_q .$$
\end{prop}
Thus any rotation $T\in \mathbf{SO}(4)$ may be represented by $T(x)=qxp^{*},$ where $q,\,p\in \s^3.$
\begin{prop}\hspace{-5pt}{\bf .}
Let be $\mathbf{SO}(4)\ni T=\Theta_{q,\,q^{*}}=T_q$. Then $T\in \mathbf{SO}(3).$
\end{prop}
Obviously, $$T(e_0)=\Theta_{q,\,q^{*}}(e_0)=T_q(e_0)=qe_0q^{*}=qq^{*}=1=e_0,$$ i.e. $T$ fixes the North Pole $e_0$ and
hence $T\in\mathbf{SO}(3).$
\begin{rem} \hspace{-5pt}{\bf .}
It is easily seen that $\s^3$ is a double covering of $\mathbf{SO}(3)$ and $\s^3\times\s^3$ is a double covering of $\mathbf{SO}(4).$
\end{rem}

Representing rotations by quaternions yields an instructive and geometrically appealing clarification of the geometry of rotations. Here we summarize the major results of Meister and Schaeben (2004) and add some explicit relationships which will prove helpful for our purposes.

\subsection{Geometrical objects of $\s^3 \subset \hz$ and of $\s^2 \subset \rz^3$}
\begin{definition} \hspace{-5pt}{\bf .}
Let $q_1, q_2 \in \s^3$ be two orthogonal unit quaternions. The set of quaternions
\begin{displaymath} \label{geoqcircle}
q(t) = q_1 \cos t + q_2 \sin t, \enspace t \in [0, 2\pi) ,
\end{displaymath}
constitutes the great--circle denoted $C(q_1,q_2) \subset \s^3$.
\end{definition}
With
\begin{equation}
q_1 = \frac{1-rh}{\Vert 1-rh \Vert} \mbox{   and    } q_2 = \frac{h + r}{\Vert h+r \Vert} \label{q1q2}
\end{equation}
for ${\bm h}, {\bm r} \in \s^2$ with ${\bm r} \neq -{\bm h}$ it is obviously
\begin{displaymath}
\mbox{Sc}(q(t)) = \cos \frac{\omega(t)}{2} = \cos \frac{\eta}{2} \cos t.
\end{displaymath}

Obviously, $({\bm h}, {\bm r})$ and $(-{\bm h}, -{\bm r})$ define the same great circle $C(q_1,q_2) \subset \s^3$.

\begin{definition} \hspace{-5pt}{\bf .}
Let $q_1, q_2, q_3, q_4 \in \s^3$ be four mutually orthonormal quaternions; let $C(q_1, q_2)$ denote the circle spanned by quaternions $q_1, q_2$, and $C(q_3, q_4)$ the orthogonal circle spanned by $q_3, q_4$. The set of quaternions
\begin{eqnarray} \label{torus}
q(s,t; \Theta) & = & \Bigl( q_1 \cos s + q_2 \sin s \Bigr) \cos \Theta + \Bigl( q_3 \cos t + q_4 \sin t \Bigr) \sin \Theta \, , \nonumber \\
& & s,t \in [0, 2\pi), \enspace \Theta \in [0, \pi/2]
\end{eqnarray}
constitutes the spherical torus denoted $T(C(q_1,q_2); \Theta) \subset \s^3$ with core $C(q_1, q_2)$.
\end{definition}

\begin{definition} \hspace{-5pt}{\bf .}
Let ${\bm r} \in \s^2$ be a unit vector and
\begin{displaymath} \label{rotaboutr}
r(t) = \cos \frac{t}{2} + {\bm r} \sin \frac{t}{2}, \enspace t \in [0, 2\pi),
\end{displaymath}
represent the rotation about ${\bm r}$ by $t \in [0, 2\pi)$. Then the set of unit vectors
\begin{equation} \label{rr_scircle}
{\bm r}'(t) = r(t) \, {\bm r}_0' \, r^{\ast}(t), \enspace t \in [0, 2\pi),
\end{equation}
with ${\bm r}_0' \in \s^2$ in the plane spanned by ${\bm h}$ and ${\bm r}$ such that ${\bm r} \cdot {\bm r}_0' = \cos (\rho), {\bm h} \cdot {\bm r}_0' = \cos (\eta-\rho)$ constitutes the small circle or cone
\begin{displaymath} \label{r_scircle}
c({\bm r}; \rho) = \{ {\bm r}' \in \s^2 \, \vert \, {\bm r} \cdot {\bm r}' = \cos \rho \} \subset \s^2
\end{displaymath}
with angle $\rho$ with respect to its centre ${\bm r}$.
\end{definition}
Its parametrized form explicitly reads (e.g. Altmann, 1986)
\begin{displaymath} \label{rotaboutr_param}
{\bm r}'(t) = {\bm r}_0' \, \cos t + ({\bm r} \times {\bm r}_0') \, \sin t + ({\bm r} \cdot {\bm r}_0') \, {\bm r} \; (1 - \cos t), \enspace t \in [0, 2\pi).
\end{displaymath}

With ${\bm h} \in \s^2$ and
\begin{displaymath} \label{rotabouth}
h(t) = \cos \frac{t}{2} + {\bm h} \sin \frac{t}{2}, \enspace t \in [0, 2\pi),
\end{displaymath}
representing the rotation about ${\bm h}$ by $t \in [0, 2\pi)$ and with ${\bm h}_0' \in \s^2$ in the plane spanned by ${\bm h}$ and ${\bm r}$ such that ${\bm h} \cdot {\bm h}_0' = \cos (\rho), {\bm r} \cdot {\bm h}_0' = \cos (\eta+\rho)$ analogously results in
\begin{equation} \label{hh_scircle}
{\bm h}'(t) = h(t) \, {\bm h}_0' \, h^{\ast}(t), \enspace t \in [0, 2\pi).
\end{equation}

\subsection{Fibres}
The fibre $G({\bm h}, {\bm r}) \subset \mathbf{SO}(3)$ of all rotations with ${\g} \, {\bm h} = {\bm r}$ is represented by the circle $C(q_1, q_2) \subset \s^3$ spanned by unit quaternions $q_1, q_2 \in \s^3$ given in terms of ${\bm h}, {\bm r} \in \s^2$ by Eqs.~(\ref{q1q2}), for example.
Therefore, the notation $C_{{\bm h}, {\bm r}} \equiv C(q_1, q_2)$ is used where it is more instructive keeping in mind that $C_{{\bm h}, {\bm r}} \equiv C_{-{\bm h}, -{\bm r}}$. Thus the major property of the circle $C(q_1, q_2)$ is that it consists of all quaternions $q(t), t \in [0, 2\pi)$, with $q(t) \, {\bm h} \, q^{\ast}(t) = {\bm r}$ for all $t \in [0, 2\pi)$, and that it covers the fibre $G({\bm h}, {\bm r})$ twice; moreover, it is uniquely characterized by the pair $({\bm h}, {\bm r}) \in \s^2 \times \s^2$ and its antipodally symmetric $(-{\bm h}, -{\bm r})$.

The set of all rotations mapping ${\bm h}$ on the small circle $c({\bm r}; \rho)$ is equal to the set of all rotations mapping all elements of the small circle $c({\bm h}; \rho)$ onto ${\bm r}$ and represented by the spherical torus $T(C(q_1, q_2); \frac{\rho}{2}) \subset \s^3$ with core $C(q_1, q_2)$.

The distance $d$ of an arbitrary $q \in \s^3$ from the circle $C(q_1,q_2)$ is given by
\begin{displaymath} \label{dqC}
d(q, C(q_1, q_2)) = \frac{1}{2} \arccos(q {\bm h} q^{\ast} \cdot {\bm r})
\end{displaymath}
If $d(q, C(q_1, q_2) = \rho$, then $q$ and $C$ are called $\rho$-incident.

Then, the torus $T(C(q_1, q_2); \frac{\rho}{2})$ consisting of all quaternions with distance $\frac{\rho}{2}$ from $C(q_1, q_2)$ essentially consists of all circles with distance $\frac{\rho}{2}$ from $C(q_1, q_2)$ representing all rotations mapping ${\bm h}$ on $c({\bm r}; \rho)$ and mapping $c({\bm h}; \rho)$ on ${\bm r}$.

The representation of a torus can be factorized in the following
way. Let $({\bm h}, {\bm u}), ({\bm r}, {\bm v}) \in \s^2 \times
\s^2$ with ${\bm u}\cdot {\bm v}=\cos \rho$. Then $C_1 \equiv
C_{{\bm h}, {\bm u}} \subset \s^3$ and $C_2 \equiv C_{{\bm r}, {\bm
v}} \subset \s^3$ exist such that
\begin{eqnarray*}
p_1 {\bm h} p_1^{\ast} & = & {\bm r}_1 \mbox{ for all } p_1 \in C_1 \\
p_2 {\bm h}_2 p_2^{\ast} & = & {\bm r}_2 \mbox{ for all } p_2 \in C_2
\end{eqnarray*}
Then $p_2^{\ast} p_1$ maps ${\bm h}$ on the small circle with centre
${\bm r}$ and angle $\rho = \arccos({\bm u} \cdot {\bm v})$ as
\begin{equation} \label{factorizetorus}
p_2^{\ast} p_1 {\bm h} p_1^{\ast} p_2 \cdot {\bm r} = p_1 {\bm h}
p_1 \cdot p_2 {\bm r} p_2^{\ast} = {\bm u} \cdot {\bm v},
\end{equation}
i.e. $\{ p_2^{\ast} p_1 \, | \, p_i \in C_i, \enspace i=1,2 \}$
represents the torus $T(C;\rho/2)$ with core $C=C_{{\bm h}, {\bm
r}}$ and angle $\rho=\arccos({\bm u} \cdot {\bm v})$. Let $p \in
T(C_{{\bm h}, {\bm r}}; \rho/2)$, then $p {\bm h} p^{\ast} = {\bm
x}$ with ${\bm x} \cdot {\bm r} = \cos \rho$. Given ${\bm u}, {\bm
v} \in \s^2$ with ${\bm u} \cdot {\bm v} = \cos \rho$ there exists a
unique $p_2 \in \s^3$ such that $p_2 {\bm x} p_2^{\ast} = {\bm u}$
and $p_2 {\bm r} p_2^{\ast} = {\bm v}$. For any $p_1 \in C_{{\bm h},
{\bm u}}$ it is $p_1^{\ast} p_2 p {\bm h} p^{\ast} p_2^{\ast} p_1 =
{\bm h}$ implying $p = p_2^{\ast} p_1$.

Since $q(t) {\bm h} q^{\ast}(t) = {\bm r}$ implies that  $q(t) {\bm h}' q^{\ast}(t) \cdot {\bm r} = {\bm h}' \cdot q^{\ast}(t) {\bm r} q(t) = {\bm h}' {\bm h}$ for all $q(t) \in C(q_1, q_2)$, the set $\{ q(t) {\bm h}' q^{\ast}(t) \}$ represents the small circle (cone) around ${\bm r}$ with angle $t = \arccos {\bm h}{\bm h}'$. In parametrized form employing Eqs.~(\ref{rr_scircle}), (\ref{hh_scircle})
\begin{equation} \label{vut}
q(t) \, {\bm h}'(u) \, q^{\ast}(t) = {\bm r}'(u+2t), \enspace t, u \in [0, 2\pi), q(t) \in C(q_1, q_2)
\end{equation}
(Meister and Schaeben, 2005), which may be rewritten for quaternions as $q(t) = r(u+2t)$.

Then the distance of $q(t) \in C(q_1, q_2)$ from an arbitrary circle $C_1$ representing all rotations mapping ${\bm h}_1$ on ${\bm r}_1$ is given by spherical trigonometry
\begin{eqnarray}
& & d(q(t), C_1) = \frac{1}{2}\arccos \Bigl( q(t) {\bm h}_1 q^{\ast}(t) \cdot {\bm r}_1 \Bigr) \label{dqtC} \\
& = & \frac{1}{2} \arccos \Bigl( {\bm h} {\bm h}_1 + {\bm r} {\bm r}_1 + \sqrt{(1-({\bm h} {\bm h}_1)^2)(1-({\bm r} {\bm r}_1)^2)} \; \cos t \Bigr). \nonumber
\end{eqnarray}

Eventually, the set of all circles $C(p_1, p_2) \subset \s^3$ with a fixed distance $\frac{\rho}{2}$ of a given $q \in \s^3$, i.e. the set of all circles tangential to the sphere $s(q; \rho/2)$ with centre $q$ and radius $\rho/2$, is characterized by
\begin{equation} \label{ctangs1}
\frac{\rho}{2} = d \Bigl(q, C(p_1, p_2)\Bigr) = \frac{1}{2} \arccos \Bigl( q \, {\bm h} \, q^{\ast} \cdot {\bm r} \Bigr),
\end{equation}
where ${\bm r} \in \s^2$ is uniquely defined in terms of ${\bm h}$ and
 $p_1, p_2$ by ${\bm r} := p(t) \, {\bm h} \,  p^{\ast}(t)$ for all $p(t) \in C(p_1, p_2)$ and any arbitrary ${\bm h} \in \s^2$,
 i.e. each circle $C(p_1, p_2)$ represents all rotations mapping some ${\bm h} \in \s^2$ onto an element of the small circle
 $c(q {\bm h} q^{\ast}; \rho)$. Thus, for each $q \in \s^3$ and $\rho \in [0, \pi)$
\begin{eqnarray}
& & \Bigl\{ C(p_1, p_2) \, \vert \, d\Bigl(q, C(p_1, p_2)\Bigr) = \frac{\rho}{2} \Bigr\} = \bigcup_{{\bm h}  \in \s^2} \enspace \bigcup_{{\bm r} \in c(q \, {\bm h} \, q^{\ast}; \rho)} C_{{\bm h}, {\bm r}} \label{ctangs2} \\
& = & \bigcup_{{\bm h}  \in \s^2} \enspace \bigcup_{{\bm r} \in c(q \, {\bm h} \, q^{\ast}; \rho)} C(p_1({\bm h},{\bm r}),p_2({\bm h}, {\bm r})) \nonumber \\
& = & \bigcup_{{\bm h}  \in \s^2_+} \enspace \bigcup_{{\bm r} \in c(q \, {\bm h} \, q^{\ast}; \rho)} C(p_1({\bm h},{\bm
r}),p_2({\bm h}, {\bm r})), \nonumber
\end{eqnarray}
where $\s^2_+$ denotes the upper hemisphere of $\s^2.$ The last equation is due to the fact that $( {\bm h}, {\bm r} )$ and $(-{\bm h}, -{\bm r})$ characterize the same great circle $C_{{\bm h}, {\bm r}} \equiv C_{-{\bm h}, -{\bm r}}$.

\section{Radon transforms}

\subsection{The spherical Radon and generalized spherical Radon transform}
Experimentally accessible crystallographic pole figures $P({\bm h}, {\bm r})$ of recorded diffracted x--ray or neutron beam intensities are modelled by
\begin{equation} \label{pdf}
P({\bm h}, {\bm r}) = \frac{1}{4\pi} \int_{G({\bm h}, {\bm r}) \cup G(-{\bm h}, {\bm r})} f({\g}) \, d{\g} ,
\end{equation}
and interpreted as the probability density function that the crystallographic axis $\pm {\bm h}$ statistically coincides with the specimen direction ${\bm r}$ given the orientation probability density function $f: \mathbf{SO}(3) \mapsto \rz_{+}^1$.

Let ${\mathcal C}$ denote the set of all $1$--dimensional totally geodesic submanifolds $C \subset \s^3$. Each $C \in {\mathcal C}$ is a $1$--sphere, i.e. a circle with centre ${\mathcal O}$. Each circle is characterized by a unique pair of unit vectors $({\bm h}, {\bm r}) \in \s^2 \times \s^2$ (and equivalently by its antipodally symmetric) by virtue of $q \, {\bm h} \, q^{\ast} = {\bm r}$ for all $q \in C$.
Thus, refering to the quaternion representation, Eq.~(\ref{pdf}) can be rewritten in a parametric form as
\begin{displaymath} \label{srt}
P({\bm h}, {\bm r}) = \frac{1}{8\pi} \int_{C \cup C^{\bot}} f(q) \, d\omega_1(q),
\end{displaymath}
where the circle $C \subset S^3$ represents all rotations mapping ${\bm h} \in \s^2$ on ${\bm r} \in \s^2$ and where $C^{\bot}(q_1, q_2)$ is the orthogonal circle representing all rotations mapping $-{\bm h}$ on ${\bm r}$, and where $\omega_1$ denotes the usual one--dimensional circular Riemann measure.

Following Helgason (1994; 1999),
\begin{definition} \hspace{-5pt}{\bf .}
\begin{displaymath}
\frac{1}{2\pi} \int_{C} f(q) \, d\omega_1(q) = \int_{C} f(q) \, dm(q) = ({\mathcal R} f)(C)
\end{displaymath}
with the normalized measure $m=\frac{1}{2\pi} \omega_1$ is referred to as $1$--dimensional spherical (totally geodesic) Radon transform of $f$ whenever $f$ is integrable on each great circle.
\end{definition}
The Radon transform of $f$ may be represented as the convolution of $f$ with the indicator function of the great circle $C$. It associates with the function $f$ its mean values over great circles $C \in {\mathcal C}$. Since each great circle is uniquely characterized by a pair $({\bm h}, {\bm r}) \in \s^2 \times \s^2$ and its antipodally symmetric, we also use the notation $({\mathcal R} f)({\bm h}, {\bm r}) \equiv ({\mathcal R} f)(-{\bm h}, -{\bm r})$ whenever it is more instructive.

It should be noted that no distinction has been made whether $f$ refers to $\mathbf{SO}(3)$ or $\s^3$, even though the form of $f$ depends on the representation of ${\g}$; in particular, with respect to $\s^3$ only even functions $f$ could be orientation probability density functions as $q \in \s^3$ and $-q$ represent the same orientation.
Then
\begin{equation} \label{Friedel1}
P({\bm h}, {\bm r}) = \frac{1}{4} \Bigl( ({\mathcal R}f)(C_{{\bm h}, {\bm r}}) + ({\mathcal R}f)(C_{-{\bm h}, {\bm r}}) \Bigr) = ({\mathcal X}f)( {\bm h}, {\bm r}),
\end{equation}
where ${\mathcal X}f$ is referred to as basic {\em crystallographic} x--ray transform.

Further following Helgason (1994; 1999) the generalized $1$--dimensional spherical Radon transform and the respective dual is well defined.
\begin{definition} \hspace{-5pt}{\bf .}
The generalized $1$--dimensional spherical Radon transform of a real function $f: \s^3 \mapsto \rz^1$ is defined as
\begin{displaymath} \label{gsrt}
({\mathcal R}^{(\rho)} f)(C) = \frac{1}{4\pi^2 \sin \rho} \, \int_{d(q, C) = \rho} \; f(q) \, dq .
\end{displaymath}
\end{definition}
The generalized Radon transform of $f$ may be represented as the convolution of $f$ with the indicator function of the torus $T$. It associates with $f$ its mean values over the torus $T(C, \rho)$ with core $C$ and radius $\rho$, Eq.~(\ref{torus}).

Following Berens et al. (1968), Freeden et al. (1998, p. 64),
\begin{definition} \hspace{-5pt}{\bf .}
The spherically generalized translation of a real function $F \in C(\s^2)$ or $F \in L^p(\s^2), 1 \le p < \infty$, is defined as
\begin{displaymath} \label{gst1}
({\mathcal T}^{(\rho)} F) ({\bm r}) = \frac{1}{2\pi \sqrt{1-\cos^2 \rho}} \int_{{\bm r} {\bm r}' = \cos \rho} F({\bm r}') d{\bm r}' .
\end{displaymath}
where $2\pi \sqrt{1-\cos^2 \rho}$ is the length of the circle $c({\bm r}; \rho)$ centered at ${\bm r}$ with radius $\cos \rho$.
\end{definition}
It can be determined by
\begin{displaymath} \label{gst2}
({\mathcal T}^{(\rho)} F) ({\bm r}) = \frac{1}{2\pi \sqrt{1-\cos^2 \rho}} \int_0^{2\pi} F({\bm r}'(t)) dt,
\end{displaymath}
with ${\bm r}'(t)$ given according to Eq.~(\ref{rr_scircle}).

When the translation ${\mathcal T}^{(\rho)}$ is applied to the Radon transform with respect to one of its arguments, then the geometry of rotations represented by quaternions amounts to
\begin{eqnarray}
\Bigl({\mathcal T}^{(\rho)} [{\mathcal R} f] \Bigr) ({\bm h}, {\bm r})
& = & \frac {1}{2\pi \sin \rho} \int_{c({\bm h}; \rho)} ({\mathcal R} f) ({\bm h}', {\bm r}) d{\bm h}' \nonumber
\\[0.5ex]
& = & \frac {1}{2\pi \sin \rho} \int_{c({\bm r}; \rho)} \; ({\mathcal R} f) ({\bm h}, {\bm r}') d{\bm r}' \label{amvt} \\[0.5ex]
& = & \frac {1}{4\pi^2 \sin \rho} \int_{c({\bm r}; \rho)} \int_{ C(\, q_1({\bm h}, {\bm r}'), q_2({\bm h}, {\bm r}') \, )} \; f(q) d\omega_1(q)
\, d{\bm r}' \nonumber \\[0.5ex]
& = & \frac{1}{4\pi^2 \sin \rho} \int_{T( C(q_1({\bm h}, {\bm r}), q_2({\bm h}, {\bm r})) ; \frac{\rho}{2})} \; f(q) dq \label{Helgason} \\[0.5ex]
& = & \frac{1}{4\pi^2 \sin \rho} \int_{d( \, q, \, C(q_1({\bm h}, {\bm r}), q_2({\bm h}, {\bm r})) \,  ) = \frac{\rho}{2}} \; f(q) dq \nonumber \\[1ex]
& = & ({\mathcal R}^{(\rho/2)} f)(C_{{\bm h}, {\bm r}}).
\end{eqnarray}
Eq.~(\ref{amvt}), cf. (Bunge, 1969, p. 47; Bunge, 1982, p. 76), is an \'Asgeirsson--type mean value theorem (cf. \'Asgeirsson, 1937; John, 1938) justifying the application of ${\mathcal T}^{(\rho)}$ to ${\mathcal R} f$ regardless of the order of its arguments, and Eq.~(\ref{Helgason}) is instrumental to the inversion of the spherical Radon transform Helgason, 1994; 1999). We have just accomplished

\begin{prop} \label{prop1} \hspace{-5pt}{\bf .}
The generalized $1$--dimensional spherical Radon transform is equal to the translated spherical Radon transform
\begin{displaymath} \label{theo1}
\Bigl({\mathcal T}^{(\rho)} [{\mathcal R} f] \Bigr) ({\bm h}, {\bm r}) = ({\mathcal R}^{(\rho/2)} f)(C_{{\bm h}, {\bm r}})
\end{displaymath}
and it can be identified with the angle density function
\begin{equation} \label{adf1}
({\mathcal A}f)({\bm h}, {\bm r}; \rho) := \frac {1}{2\pi \sin \rho} \int_{c({\bm r}; \rho)} \; ({\mathcal R} f) ({\bm h}, {\bm r}') d{\bm r}'
\end{equation}
\end{prop}
The angle density function $({\mathcal A}f)({\bm h}, {\bm r}; \rho)$ has been introduced in (Bunge, 1969, p. 44; Bunge, 1982, p. 74) (with a false normalization). It is the probability density that the crystallographic direction ${\bm h}$ statistically encloses the angle $\rho, 0 \leq \rho \leq \pi,$ with the specimen direction ${\bm r}$ given the orientation probability density function $f$.
It should be noted that
\begin{eqnarray}
({\mathcal A}f)({\bm h}, {\bm r}; 0) & = & ({\mathcal R} f) ({\bm h}, {\bm r}) , \label{ar}\\
({\mathcal A}f)({\bm h}, {\bm r}; \pi) & = & ({\mathcal R} f) ({\bm h}, -{\bm r}) . \nonumber
\end{eqnarray}

Finally, with respect to the diffraction experiment, it should be
noticed that
\begin{align}
\left({\mathcal T}^{\rho}[{\mathcal X}f]\right)({\bm h},\,{\bm r}) &
= \frac{1}{4} \left( ({\mathcal R}^{(\rho /2)}f)(C)+
({\mathcal R}^{(\rho /2)}f)(C^{\bot}) \right) \nonumber \\
 & \frac{1}{4}\left( ({\mathcal A}f)({\bm h},\,{\bm r};\,\rho) + ({\mathcal A}f)(-{\bm h},\,{\bm
 r};\,\rho)\right) \nonumber \\
  & = ({\mathcal W})({\bm h},\,{\bm r};\,\rho)
\end{align}
Thus with respect to a diffraction experiment, $({\mathcal W})({\bm
h},\,{\bm r};\,\rho)$ is at once accessible if the specimen rapidly
rotates around the specimen direction ${\bm r}$ during the
measurements.

\subsection{Kernels and their twofold Radon transform}
Now let $K$ be a kernel function $S^3 \times S^3 \mapsto \rz_{+}^1$. Then we may apply the Radon transform twice in the sense that we apply it once with respect to the first and once with respect to the second variable (cf. Boogaart et al., 2005), i.e.
\begin{displaymath}
({\mathcal R} [K(\circ_1, p_2 )])(C_1) = \frac{1}{2\pi} \int_{C_1} \, K(p_1, p_2) \, d\omega_1(p_1) = F(C_1, p_2)
\end{displaymath}
and
\begin{eqnarray*}
({\mathcal R} [({\mathcal R} [K(\circ_1, \circ_2 )])(C_1)]) (C_2) = ({\mathcal R} [F(C_1, \circ_2)]) (C_2) \\
 = \frac{1}{4\pi^2} \int_{C_2} \int_{C_1} \, K(p_1, p_2) \, d\omega_1(p_1) d\omega_1(p_2) = G(C_1, C_2),
\end{eqnarray*}
where $C_1$ denotes the great circle representing all rotations mapping ${\bm h}_1$ on ${\bm r}_1$, and analogously $C_2$ with respect to ${\bm h}_2$ and ${\bm r}_2$.

In case the kernel $K$ depends only on the quaternion product $p_1^{\ast} p_2$, then
\begin{eqnarray*}
({\mathcal R} [({\mathcal R} [K(\circ_1^{\ast} \circ_2 )])(C_1)]) (C_2) & = & \frac{1}{4 \pi^2} \int_{p_1 {\bm h}_1 p_1^{\ast} = {\bm r}_1} \int_{p_2 {\bm h}_2 p_2^{\ast} = {\bm r}_2} \, K(p_1^{\ast} p_2) \, d\omega_1(p_1) \, d\omega_1(p_2) \\
& = &  \frac{1}{4 \pi^2 \sqrt{1-({\bm r}_1 \cdot {\bm r}_2)^2}} \int_{(p {\bm h}_1 p^{\ast} \cdot {\bm h}_2) = ({\bm r}_1 \cdot {\bm r}_2) } \, K(p) \, dp \\
& = & \frac{1}{4 \pi^2 \sin (\rho/2)} \int_{T(C; \rho/2)} \, K(p) \, dp, \\
& = & ({\mathcal R}^{(\rho/2)} K) (C) \\
& = & ({\mathcal A}K)({\bm h}_1, {\bm h}_2; \rho)
\end{eqnarray*}
with the core $C = C_{{\bm h}_1, {\bm h}_2}$ and $\rho = \arccos({\bm r}_1 \cdot {\bm r}_2)$ according to Eq.~(\ref{factorizetorus}).

In case the kernel $K$ depends only on  $\omega/2 = \arccos (\mbox{Sc}(p_1^{\ast} p_2))$, then the function $F$ depends only on $\arccos (p_2 {\bm h}_1 p_2^{\ast} \cdot {\bm r}_1) = d(p_2, C_1)$. As a function of the variable $p_2$ with parameters ${\bm h}_1, {\bm r}_1$ implicitly provided by $C_1$, $F$ may be referred to as fibre function. They share most of the characteristics of ridge functions initially defined in linear spaces and discussed in (Donoho, 2000). Then the function $G$ depends only on $\arccos({\bm h}_1 {\bm h}_2)$ and $\arccos({\bm r}_1 {\bm r}_2)$, cf. (Boogaart et al., 2005).


\begin{prop} \label{prop2} \hspace{-5pt}{\bf .}
For a kernel function of the form $K(p_1, p_2) = K(p_1^{\ast} p_2)$, its angle density function, i.e. its generalized spherical Radon transform, is identical with the twofold application of the spherical Radon transform with respect to the two components of its argument.
\end{prop}

\subsection{The dual spherical Radon and dual generalized spherical Radon transform}
Following again Helgason (1994, 1999) we have two more definitions.
\begin{definition} \hspace{-5pt}{\bf .}
The dual $1$--dimensional spherical Radon transform of a real continuous function $\varphi: {\mathcal C} \mapsto \rz^1$ is defined as
\begin{displaymath} \label{dsrt}
({\widetilde {\mathcal R}} \varphi) (q) = \int_{C \ni q}   \varphi(C) \, d\mu(C) ,
\end{displaymath}
where $\mu$ denotes the unique measure on the compact space $ C \in {\cal C}: q \in C $, invariant under all rotations around $q$, and having total measure $1$.
\end{definition}
Thus $({\widetilde {\mathcal R}} \varphi) (q)$ is the mean value of $\varphi$ over the set of circles $C \in {\mathcal C}$ passing through $q$.

\begin{definition} \hspace{-5pt}{\bf .}
The dual generalized $1$--dimensional spherical Radon transform of a real continuous function $\varphi: {\mathcal C} \mapsto \rz^1$ is defined as
\begin{displaymath} \label{gdsrt}
({\widetilde {\mathcal R}}^{(\rho)} \varphi) (q) = \int_{\{C \in \,\mathcal{C}:\,d(q, C) = \rho\}}   \varphi(C) \, d{\widetilde{\mu}}(C) .
\end{displaymath}
\end{definition}
With the normalized measure ${\widetilde{\mu}} = A^{-1} (\rho) d\mu, \enspace A(\rho) = 4 \pi sin^2 \rho$, it is the mean value of $\varphi$ over the set of all circles $C \in {\mathcal C}$ with distance $\rho$ from $q$.

Again, geometrical reasoning, Eqs.~(\ref{ctangs1}), (\ref{ctangs2}), yields that $ \Bigl( {\widetilde {\mathcal R}}^{(\rho)} \Bigl[ ({\mathcal R} f) (\circ) \Bigr] \Bigr) (q) $ is the mean value of $({\mathcal R} f)$ over the set of all circles $C \in {\mathcal C}$ with distance $\rho$ from $q$, i.e. tangential to the sphere $s(q; \rho)$. More specifically, with the usual two--dimensional spherical Riemann measure $\omega_2$,
\begin{align}
\Bigl({\widetilde {\mathcal R}}^{(\rho)} \Bigl[ ({\mathcal R} f) (\circ) \Bigr] \Bigr) (q)
& = \int_{d(q, C) = \rho} \, ({\mathcal R} f)(C_{{\bm h}, {\bm r}}) \, d\mu(C_{{\bm h}, {\bm r}}) \nonumber \\
& = \frac{1}{4 \pi \sin (\rho/2)} \int_{\s^2} \int_{c(q{\bm h}q^{\ast}; \rho/2)} \, ({\mathcal R} f)(C_{{\bm h}, {\bm r}}) \, d{\bm r} d\omega_2({\bm h}) \\
& = \frac{1}{2} \int_{\s^2} ({\mathcal A}f)({\bm h}, q {\bm h}
q^{\ast}; \rho/2) \, d\omega_2({\bm h}) \label{angular}, \\
& = \frac{1}{2} \int_{\s^2} ({\mathcal R}^{(\rho )}f)(C_{{\bm
h},\,q{\bm h}q^{\ast}} \, d\omega_2({\bm h}) \\
& = \left({\widetilde {\mathcal R}} \Bigl[ ({\mathcal R} f) (\circ)
\Bigr] \right) (q),
\end{align}
which is instrumental for the inversion of the spherical Radon transforms. In particular,
\begin{equation} \label{sangular}
\Bigl({\widetilde {\mathcal R}} \Bigl[ ({\mathcal R} f) (\circ) \Bigr] \Bigr) (q) = \frac{1}{2} \int_{\s^2} ({\mathcal R}f)({\bm h}, q {\bm h} q^{\ast}) \, d\omega_2({\bm h}) ,
\end{equation}

Next we want to describe the generalized dual Radon transform in a more group-theoretical way (cf. Helgason, 1994; 1999). Let $C_0\in\mathcal{C}$ be a fixed great circle with distance $\rho$ from $q,$ i.e. $d(q,\,C_0)=\rho .$ Let $\Theta_{r,\,s^{*}}$ be the rotation that maps the North Pole $e_0$ to $q,$ i.e. $q=re_0s^{*}=rs^{*}.$ Since the distance on $\s^3$ is rotational invariant we obtain
$$ d(q,\,C_0)=d(rs^{*},\,C_0)= d(e_0,\,r^{*}C_0s). $$
Since all rotations in $\mathbf{SO}(3)$ leave the North Pole $e_0$ invariant we conclude
$$ d(q,\,C_0)= d(e_0,\,r^{*}C_0s)=d(ke_0k^{*},\,kr^{*}C_0sk^{*})= d(e_0,\,kr^{*}C_0sk^{*}), $$
hence the set $\{kr^{*}C_0sk^{*}, r,\,s\in\s^3\ \mbox{fixed,}\ k\in\s^3\}$ is the set of all great circles that have distance $\rho$ from the North Pole $e_0.$ Now, the North Pole $e_0$ is rotated back into $q$ leading to
$$d(q,\,C_0)= d(e_0,\,kr^{*}C_0sk^{*})= d(re_0s^{*},\,rkr^{*}C_0sk^{*}s^{*})=d(q,\,rkr^{*}C_0sk^{*}s^{*}) .$$
Thus the set of all great circles with distance $\rho$ from $q$ is given by
$$ \{rkr^{*}C_0sk^{*}s^{*},\ k\in\s^3, q=rs^{*}\}.$$

The previous consideration amounts to a new description of the generalized dual Radon transform:
\begin{align*}
({\widetilde {\mathcal R}}^{(\rho)} \varphi) (q) & = \int_{\{C\in\,\mathcal{C}:\,d(q, C) = \rho\}}   \varphi(C) \,
d\mu(C)\\
 & = \int_{\mathbf{SO}(3)} \varphi(rkr^{*}C_0sk^{*}s^{*})\,dk,
 \end{align*}
 where $dk = \frac{1}{8\pi ^2}\sin \beta\,d\alpha\,d\beta\,d\gamma $ is the invariant Haar measure on
 $\mathbf{SO(3)},$\  $C_0\in\mathcal{C}$ a fixed great circle such that $d(q,\,C_0)=\rho $ and $q=rs^{*}.$

Analogously to the spherically generalized translation for function on $\s^2$, we have
\begin{definition} \hspace{-5pt}{\bf .}
The spherically generalized translation of a real function $f \in C(\s^3)$ or $F \in L^p(\s^3), 1 \le p < \infty$, is defined by the mean value operator
\begin{displaymath} \label{gst2}
\Bigl( {\mathcal T}^{(\rho)} f \Bigr) (q) = \frac{1}{A(\rho)} \int_{q^{\ast} p = \cos \rho} f(p) dp ,
\end{displaymath}
where $A(\rho)$ denotes the surface area of the sphere $s(q;\rho)$ centered at $q \in \s^3$ with radius $\cos \rho$.
\end{definition}

Then the following proposition can be shown.
\begin{prop} \label{prop3} \hspace{-5pt}{\bf .}
\begin{eqnarray}
\Bigl({\widetilde {\mathcal R}}^{(\rho)} \Bigl[ ({\mathcal R} f) (\circ) \Bigr] \Bigr) (q)
& = & \int_{\{C\in\,\mathcal{C}:\,d(q, C) = \rho\}} ({\mathcal R} f)(C) \, d{\widetilde{\mu}}(C) \nonumber \\
& = & \int_{C_0} ({\mathcal T}^{d(q,p)} f) (q) \, dm(p), \label{st}
\end{eqnarray}
where $C_0$ is a fixed great circle with distance $d(q,\,C_0)=\rho$ from $q.$
\end{prop}

\begin{eqnarray*}
\Bigl({\widetilde {\mathcal R}}^{(\rho)} \Bigl[ ({\mathcal R} f) (\circ) \Bigr] \Bigr) (q) & = &
\int_{\{C\in\mathcal{C}:\,d(q, C) = \rho\}}  ({\mathcal R} f)(C) \, d{\widetilde{\mu}}(C) \\
 & = & \int_{\mathbf{SO}(3)} ({\mathcal R} f)(rkr^{*}C_0sk^{*}s^{*})\,dk \\
 & = & \int_{\mathbf{SO}(3)} \int_{C_0} f(rkr^{*}psk^{*}s^{*})\,dm(p)\,dk \\
 & = & \int_{C_0}\int_{\mathbf{SO}(3)} f(rkr^{*}psk^{*}s^{*})\,dk\,dm(p)
\end{eqnarray*}
While $k$ varies in $\mathbf{SO}(3)$, the set $\{rkrpsk^{*}s^{*},\ k\in\mathbf{SO}(3)\}$ is a small sphere around $q$ containing $p:$
\begin{eqnarray*}
d(q,\,rkrpsk^{*}s^{*})=d(rs^{*},\,rkrpsk^{*}s^{*})=d(e_0,\,krpsk^{*})\\
=d(k^{*}e_0k,\,r^{*}ps)=d(e_0,\,r^{*}ps)=d(rs^{*},\,p)=d(q,\,p) \enspace ,
\end{eqnarray*}
and $dk = A^{-1} (\rho)\,dp$ is the normalized Riemann measure of the small sphere $s(q;\,\rho)$ with center $q$ and radius $\rho=d(q,p).$
Hence
\begin{eqnarray}
\Bigl({\widetilde {\mathcal R}}^{(\rho)} \Bigl[ ({\mathcal R} f) (\circ) \Bigr] \Bigr) (q) & = &
\int_{C_0}\int_{\mathbf{SO}(3)} f(rkr^{*}psk^{*}s^{*})\,dk\,dm(p) \nonumber \\
& = & \int_{C_0} \frac{1}{A(\rho)}\int_{S_{\rho}(q)} f(p)\,dp\,dm(p) \nonumber \\
& = & \int_{C_0} (\mathcal{T}^{d(q,\,p)}f)(q)\,dm(p) \label{doppeldach} .
\end{eqnarray}

\begin{rem} \hspace{-5pt}{\bf .}
A special situation occurs when $f$ is a central function. Assuming that $f$ is central with respect to $q_0$,
$({\mathcal R} f)$ is constant on the set of all great circles $C \in {\mathcal C}$ for which $q_0 {\bm h}_C q_0^{\ast}
\cdot {\bm r}_C$, where ${\bm h}_C, {\bm r}_C$ denote the unit vectors associated to the great circle $C$. This is true
for the set of all great circles with $d(q_0, C) = \rho$ as they are characterized by Eq.~(\ref{ctangs1}). Thus we can
drop the outer integration and find for the right hand side of Eq.~(\ref{st})
\begin{displaymath}
\Bigl({\widetilde {\mathcal R}}^{(\rho)} \Bigl[ ({\mathcal R} f) (\circ) \Bigr] \Bigr) (q_0) = ({\mathcal R} f)(C) \mbox{ for some $C$ with } d(q_0, C) = \rho .
\end{displaymath}
Moreover, in this case, $({\mathcal T}^{\rho} f) (q_0) \equiv f(q_0)$ for all $\rho$.  Thus the left hand side of Eq.~(\ref{st})
\begin{displaymath}
\int_{d(q_0,C) = \rho} ({\mathcal T}^{d(q_0,p)} f) (q_0) \, dm(p) = ({\mathcal R} f)(C) \mbox{ for some $C$ with } d(q_0, C) = \rho .
\end{displaymath}
\end{rem}

Parameterizing the great circle $C_0$ by polar coordinates and assume $f$ even $(f(q)=f(-q))$ Eq.~(\ref{doppeldach})
takes the form
\begin{eqnarray}
\Bigl({\widetilde {\mathcal R}}^{(\rho)} \Bigl[ ({\mathcal R} f) (\circ) \Bigr] \Bigr) (q) & = & 4
\int_0^{\frac{\pi}{2}} (\mathcal{T}^{d(q,\,p)}f)(q) \,d\tau \label{doppeldach1}
\end{eqnarray}
Denote by $q_0\in C_0$ a point with minimum distance from $q,$ i.e. $d(q_0,q)=\rho$. Using spherical trigonometry with respect to the triangle $qq_0p$ we obtain
$$ \cos d(q,p) = \cos d(q_0,q) \,\cos d(q_0,\,p) .$$
Fix $q$ and set $v= \cos d(q_0,q)$ and $u=v\,\cos d(q_0,\,p)= \cos d(q,p),$
$$ F(u) = (\mathcal{T}^{d(q,\,p)}f)(q), \quad \widehat{F}(v)= \Bigl({\widetilde {\mathcal R}}^{(\rho)} \Bigl[ ({\mathcal R} f) (\circ) \Bigr] \Bigr)
(q).$$ Then Eq.~(\ref{doppeldach1}) becomes an Abel's integral equation
$$ \widehat{F}(v) = 4\int_0^v \frac{F(u)}{\sqrt{v^2-u^2}}\,du, $$
which is inverted by
$$ F(u)=\frac{1}{2\pi} \frac{d}{du} \int_0^u \widehat{F}(v)\,\frac{v}{\sqrt{u^2-v^2}}\,dv .$$
Since $ F(1)=\lim_{u\to 1}(\mathcal{T}^{\arccos u}f)(q)=f(q),$ we get the following inversion formula (cf. Helgason
(1999))
$$ f(q)= \frac{1}{2\pi} \left[ \frac{d}{du} \int_0^u \Bigl({\widetilde {\mathcal R}}^{(\arccos v)} \Bigl[ ({\mathcal R} f) (\circ) \Bigr]
\Bigr)(q)\frac{v}{\sqrt{u^2-v^2}}\,dv \right]_{u=1} ,$$
and with Eq.~(\ref{angular}) we obtain
$$ f(q)= \frac{1}{4\pi} \left[ \frac{d}{du} \int_0^u \int_{\s^2} ({\mathcal A}f)({\bm h}, q {\bm h} q^{\ast}; 2\arccos v) \, d\omega_2({\bm h}) \frac{v}{\sqrt{u^2-v^2}}\,dv \right]_{u=1} ,$$
which will be further transformed. First, we substitute $t=v^2$, getting
\begin{eqnarray*}
\frac{1}{8\pi}\,\frac{d}{du}\int_0^{u^2} \int_{\s^2} ({\mathcal A}f)({\bm h}, q
{\bm h} q^{\ast}; 2\arccos \sqrt{t}) \, d\omega_2({\bm h}) \frac{1}{\sqrt{u^2-t}}\,dt \Big|_{u=1} \\
= \frac{1}{4\pi}\,\frac{d}{du^2}\int_0^{u^2} \int_{\s^2}({\mathcal A}f)({\bm h}, q {\bm h} q^{\ast}; 2\arccos \sqrt{t})
\, d\omega_2({\bm h}) \frac{1}{\sqrt{u^2-t}}\,dt \Big|_{u=1}.
\end{eqnarray*}
Next, we put $s=u^2$ to get
$$ = \frac{1}{4\pi}\,\frac{d}{ds}\int_0^{s}\int_{\s^2} ({\mathcal A}f)({\bm h}, q {\bm h} q^{\ast}; 2\arccos \sqrt{t}) \, d\omega_2({\bm h}) \frac{1}{\sqrt{s-t}}\,dt \Big|_{s=1}.$$ To shift the singularity, we set $w=s-t$ which gives
$$= \frac{1}{4\pi}\,\frac{d}{ds}\int_0^{s}\int_{\s^2} ({\mathcal A}f)({\bm h}, q {\bm h} q^{\ast}; 2\arccos
\sqrt{s-w}) \, d\omega_2({\bm h}) \frac{1}{\sqrt{w}}\,dw \Big|_{s=1}. $$
Now, differentiation  gives
\begin{multline*} =
\frac{1}{4\pi} \left( \int_{\s^2} ({\mathcal A}f)({\bm h}, q {\bm h} q^{\ast}; \pi) \, d\omega_2({\bm h}) \right. \\
\left. + \int_0^{s}\int_{\s^2} \frac{d}{ds}({\mathcal A}f)({\bm h}, q {\bm h} q^{\ast}; 2\arccos \sqrt{s-w}) \, d{\bm
h} \frac{1}{\sqrt{w}}\,dw \right)\Big|_{s=1}
\end{multline*}
Using $\frac{d}{ds}({\mathcal A}f)=-\frac{d}{dw}({\mathcal A}f)$ and taking into account $({\mathcal A}f)({\bm h},
q{\bm h} q^{\ast}; \pi)=({\mathcal R}f)({\bm h}, -q {\bm h} q^{\ast})$ (cf. Eq.~(\ref{ar})) and $s=1$ result in
\begin{multline*} =
\frac{1}{4\pi} \left( \int_{\s^2} ({\mathcal R}f)({\bm h}, -q {\bm h} q^{\ast}) \, d\omega_2({\bm h}) \right. \\
\left. - \int_0^{1}\int_{\s^2} \frac{d}{dw}\left(({\mathcal A}f)({\bm h}, q {\bm h} q^{\ast}; 2\arccos
\sqrt{1-w})\right) \, d\omega_2({\bm h}) \frac{1}{\sqrt{w}}\,dw \right)
\end{multline*}
and finally put $2w=1-\cos \theta$ to obtain
\begin{multline*}
 =\frac{1}{4\pi} \left( \int_{\s^2} ({\mathcal R}f)({\bm h}, -q {\bm h} q^{\ast}) \, d\omega_2({\bm h}) \right. \\
\left. + 2 \int_0^{\pi}\int_{\s^2} \frac{d}{d\cos \theta}\left(({\mathcal A}f)({\bm h}, q {\bm h} q^{\ast}; \theta)\right) \,
 d\omega_2({\bm h}) \cos\tfrac{\theta}{2}\,d\theta \right).
\end{multline*}


\section{Inversion of the generalized spherical Radon transform}
For $f \in C^{\infty}(\s^3)$ the spherical Radon transformation ${\mathcal R}: f \mapsto {\mathcal R} f$ has a kernel consisting of the odd functions, i.e. the functions satisfying $f(q) + f(-q) = 0$. Since $q$ and $-q$ represent the same rotation, and since $f$ is a probability density function, we are interested in even functions $f: \s^3 \mapsto \rz_{+}^1$ only. In practical texture analysis, we are especially interested to recover $f$ from data $w_i = w({\bm h}_i, {\bm r}_i; \rho_i)$ originating in sampling $({\mathcal W}f)({\bm h}, {\bm r}; \rho)$ for discrete values of ${\bm h}, {\bm r} \in \s^2$ and $\rho \in (0, \pi/2)$. Obviously, due to the additional symmetry Eqs.~(\ref{Friedel1}), \ref{Friedel2}), introduced by Friedel's law, $f$ cannot completely be recovered from integral intensity data $\iota_i = P({\bm h}_i, {\bm r}_i)$ nor by mean integral intensity data $w_i = w({\bm h}_i, {\bm r}_i; \rho_i)$. In terms of spherical harmonics, only harmonic coeffficients with respect to harmonics of even order can be determined from diffraction intensity data (Matthies, 1979).

\subsection{Inversion of the spherical Radon transform}
In texture analysis, i.e. in material science and engineering, the best known inversion formula dates back right to the begining of ``quantitative'' texture anaylsis (Bunge, 1965; Roe, 1965). The formula may be rewritten in a rather abstract way as
\begin{equation} \label{harm}
f  = {\mathcal F}_{SO(3)}^{-1} \, {\mathcal S} \, {\mathcal F}_{\s^2 \times \s^2} \, {\mathcal R}f \enspace ,
\end{equation}
where ${\mathcal S}$ denotes a scaling matrix with entries ${\sqrt{2\ell +1}}$ indicating the ill-posedness of the inverse problem. It states that the harmonic coefficients of $f$ are up to a scaling equal to the harmonic coefficiens of its Radon transform. In the context of texture analysis and experimentally accessible ``pole figures'', the former statement is true only for even--order coefficients.

The first analytical inversion formula
was conributed by Matthies (1979) and then rewritten in terms of the angle density fucntion
(Muller et al., 1981).

In integral geometry the following inversion formlae are known
\begin{eqnarray}
f & = & \frac{1}{4 \pi} \int_{S^2} (-2\Delta_{\s^2 \times \s^2} +1)^{1/2}({\mathcal R}f)({\bm h}, \, {\g} {\bm h})\, d\omega_2({\bm h}) \nonumber \\
  & = & \frac{1}{2 \pi} {\widetilde {\mathcal R}} \, [(-2\Delta_{\s^2 \times \s^2} +1)^{1/2} \, ({\mathcal R}f)]  \label{i1}
\end{eqnarray}
where $\widetilde{{\mathcal R}}$ denotes the adjoint operator of
${\mathcal R}$ with respect to $L^2$, and with respect to
Eq.~(\ref{sangular})
\begin{eqnarray}
f & = & \frac{1}{2 \pi} (-4\Delta_{\s^3} +1)^{1/2} \, {\widetilde {\mathcal R}} \, {\mathcal R}f \nonumber \\
  & = & \frac{1}{2 \pi} \Bigl( (-4\Delta_{\s^3} +1)^{1/2} \, {\widetilde {\mathcal R}} \Bigr) \; {\mathcal R}f \label{i21}\\
  & = & \frac{1}{2 \pi} (-4\Delta_{\s^3} +1)^{1/2} \, \Bigl( {\widetilde {\mathcal R}} {\mathcal R}f \Bigr) \label{i22}
\end{eqnarray}
cf. (Helgason, 1994; 1999). The equivalence of all these formula is shown in (Bernstein and Schaeben, 2005).

Comparing Eq.~(\ref{i21}) with Eq.~(\ref{i22}) with respect to Eq.~(\ref{harm}) we may ask ourselves:
Is
\begin{displaymath}
(-4\Delta_{\s^3} +1)^{1/2} \, \Bigl( {\widetilde {\mathcal R}} {\mathcal R}f \Bigr)
\end{displaymath}
a better conditioned inverse problem in practical applications than
\begin{displaymath}
\Bigl( (-4\Delta_{\s^3} +1)^{1/2} \, {\widetilde {\mathcal R}} \Bigr) \; {\mathcal R}f
\end{displaymath}
which is conditioned like
\begin{displaymath}
2 \pi \, {\mathcal F}_{SO(3)}^{-1} \, {\mathcal S} \, {\mathcal F}_{\s^2 \times \s^2}
\end{displaymath}
Obviously, $(-4\Delta_{\s^3} +1)^{1/2}$ is worse conditioned than $(-4\Delta_{\s^3} +1)^{1/2} {\widetilde {\mathcal R}}$, and ${\widetilde {\mathcal R}} {\mathcal R}$ is better conditioned than ${\mathcal R}$. Thus, the question can be specified whether the differences just cancel out, or does the later improvement dominate? This problem will be pursued elsewhere.

\subsection{Harmonic series expansion}
In terms of spherical hamonics, Eqs.~(\ref{Friedel2}, \ref{adf1}) result in
\begin{eqnarray}
({\mathcal W}f)({\bm h}, {\bm r}; \rho) & = & \frac {1}{2\pi {\sqrt{1-\cos^2 \rho}}} \, \int_{c({\bm r}; \rho)} \; \sum_{\ell=2l} \sum_{m,n} C_{\ell}^{mn} Y_{\ell}^m ({\bm h}) Y_{\ell}^n ({\bm r}') d{\bm r}' \nonumber \\ & = & \frac {1}{2\pi {\sqrt{1-\cos^2 \rho}}} \; \sum_{\ell=2l} \sum_{m,n} C_{\ell}^{mn} Y_{\ell}^m ({\bm h}) \int_{c({\bm r}; \rho)} Y_{\ell}^n ({\bm r}') d{\bm r}' \nonumber .
\end{eqnarray}
Applying
\begin{displaymath}
\frac {1}{2\pi {\sqrt{1-\tau^2}}} \, \int_{{\bm r} {\bm r}' = \tau} Y_{\ell}^n ({\bm r}') d{\bm r}' = P_{\ell} (\tau) Y_{\ell}^n ({\bm r})
\end{displaymath}
(Freeden et al., 1998, p. 64) with $\tau = \cos \rho$ finally yields
\begin{eqnarray} \label{sexpanW}
({\mathcal W}f)({\bm h}, {\bm r}; \rho) = \sum_{\ell=2l} P_{\ell}(\cos \rho) \sum_{m,n} C_{\ell}^{mn} Y_{\ell}^m ({\bm h}) Y_{\ell}^n ({\bm r}) ,
\end{eqnarray}
(cf. Bunge, 1969, p. 45; Bunge, 1982, p. 74).

In case of an orientation probability function $f(\arccos(\mbox{Sc}(q_0^{\ast} q)))$ which is radially symmtric with respect to a given $q_0$, Eq.~(\ref{sexpanW}) simplifies further to
\begin{eqnarray} \label{sexpanWrs}
({\mathcal W}f)({\bm h}, {\bm r}; \rho)
& = & \sum_{\ell=2l} P_{\ell}(\cos \rho) \sum_{m,n} C_{\ell}^{mn} Y_{\ell}^m ({\bm h}) Y_{\ell}^n ({\bm r}) \nonumber \\
& = & \sum_{\ell=2l} \frac{2 {\ell} +1}{4 \pi} C_{\ell} P_{\ell}(\cos \rho)  P_{\ell} (\cos \eta)
\end{eqnarray}
as the Radon transform becomes a function of $\eta = \arccos(q_0 {\bm h} q_0^{\ast} \cdot {\bm r}) = d(q_0, C)$ only,
where $C$ denotes the great circle representing all rotations mapping ${\bm h}$ on ${\bm r}$. It should be noted that
$\rho$ and $\eta$ commute, i.e.
\begin{displaymath}
({\mathcal W}f)({\bm h}, {\bm r}; \rho) \vert_{q_0 {\bm h} q_0^{\ast} \cdot {\bm r} = \cos \eta} = ({\mathcal W}f)({\bm h}, {\bm r}; \eta) \vert_{q_0 {\bm h} q_0^{\ast} \cdot {\bm r} = \cos \rho}
\end{displaymath}
Integrating the radially symmetric spherical Radon transform over a small circle with angle $\rho$ with respect to ${\bm r}$ with $q_0 {\bm h} q_0^{\ast} \cdot {\bm r} = \cos \eta$ is equal to integrating the spherical Radon transform over a small circle with angle $\eta$ with respect to ${\bm r}$ with $q_0 {\bm h} q_0^{\ast} \cdot {\bm r} = \cos \rho$. In terms of the generalized spherical Radon transform we find

\begin{prop} \label{prop4} \hspace{-5pt}{\bf .}
For a radially symmetric function $f$
\begin{displaymath}
({\mathcal R}^{(\rho/2)} f)(C_{{\bm h}, {\bm r}}) \vert_{q_0 {\bm h} q_0^{\ast} \cdot {\bm r} = \cos \eta}  = ({\mathcal R}^{(\eta/2)} f)(C_{{\bm h}, {\bm r}}) \vert_{q_0 {\bm h} q_0^{\ast} \cdot {\bm r} = \cos \rho}
\end{displaymath}
\end{prop}

Sampling $({\mathcal W}f)({\bm h}, {\bm r}; \rho)$ for discrete values of ${\bm h}, {\bm r} \in \s^2$ and $\rho \in (0, \pi/2)$ gives rise to data $w_i = w({\bm h}_i, {\bm r}_i; \rho_i)$ and a system of linear equations
\begin{displaymath}
w_i = \sum_{\ell=2l} P_{\ell}(\cos \rho_i) \sum_{m,n} C_{\ell}^{mn} Y_{\ell}^m ({\bm h}_i) Y_{\ell}^n ({\bm r}_i), i = 1,\ldots,I
\end{displaymath}
which may allow a solution for the $C_{\ell}^{mn}$--coefficients with even $\ell$.

\subsection{Convolution kernels and radial basis functions}
Assume that the orientation probability density function shall be modeled by the superposition of ``components'' or kernels $K_i (q)$, i.e.
\begin{displaymath}
f(q) = \sum_j \lambda_j \, K_j ({\tilde q}, q) = \sum_j \lambda_j K(q_j, q).
\end{displaymath}
Usually in practice, we choose radially symmetric kernels which give rise to
\begin{displaymath}
f(q) = \sum_j \lambda_j K(\arccos(\mbox{Sc}(q_j^{\ast} q))).
\end{displaymath}

Then the corresponding generalized spherical Radon transform ${\mathcal W}f$ is the superposition of the correspondingly generalized spherically Radon transformed kernels $k_j = {\mathcal W} K_j$
\begin{eqnarray*}
({\mathcal W}f)({\bm h}, {\bm r}; \rho) & = & \sum_j \lambda_j \, ({\mathcal W}K_j)({\bm h}, {\bm r}; \rho) \\
                                                                            & = & \sum_j \lambda_j \, k_j({\bm h}, {\bm r}; \rho) ,
\end{eqnarray*}
which may be fitted to the experimental data
\begin{displaymath}
w_i \approx \sum_j \lambda_j^{\ast} \; k({\bm h}_i, {\bm r}_i; \rho_i)
\end{displaymath}
in some sense, e.g. in the sense of a Hilbert-Sobolev norm as developed in (Boogaart et al., 2005). Then, an approximate orientation probability density function explaining the data is given by
\begin{displaymath}
f(q) = \sum_j \lambda_j^{\ast} \; K(\arccos(\mbox{Sc}(q_j^{\ast} q))) .
\end{displaymath}

\section{Examples}
In the following we provide some formulae for the Abel--Poisson (in probability: Cauchy) and the de la Vall\'ee Poussin kernel, their Radon transform and their twofold Radon transform.

In texture analysis the Abel--Poisson kernel is referred to as Lorentz standard function (Matthies et al., 1987, p. 98; Matthies et al., 1990, p. 477), and the formulae were actually taken from there. Obviously, they were initially not related to reproducing kernels and their twofold spherical Radon transform. The de la Vall\'ee Poussin kernel has been introduced into texture analysis because of its harmonic series expansion is finite (Schaeben, 1997; 1999).

For the symmetrical kernel $K(p_1, p_2) = K(p_1^{\ast} p_2)$ defined on $\s^3 \times \s^3$, the variable $\omega = 2 \arccos (\mbox{Sc}(p_1^{\ast} p_2))$ denotes the angle of the rotation of $p_1^{\ast} p_2$; for the Radon transformed kernel ${\mathcal R} K_{\s^3}(p,{\bm h},{\bm r})$ defined on $\s^3 \times \s^2 \times \s^2$, the variable $\eta$ denotes the angle $\angle(p{\bm h}p^{\ast},{\bm r})$. The two variables $\eta_1$ and $\eta_2$ of the twofold Radon transformed kernel ${\mathcal R} {\mathcal R} K({\bm h}_1,{\bm r}_1,{\bm h}_2,{\bm r}_2)$ defined on $(\s^2 \times \s^2) \times (\s^2 \times \s^2)$ correspond to the angles $\angle {\bm h}_1{\bm h}_2$ and $\angle {\bm r}_1 {\bm r}_2$. The Gegenbauer respectively the Legendre coefficients of the kernels are denoted $a_{\ell}$.

In the table we have used the following notations for special functions: B Beta function, $_2\mbox{F}_1$ hypergeometric function, and $\Gamma$ Gamma function,.

For the Abel--Poisson kernel we have
\begin{eqnarray*}
 a_{\ell} & = & (2\ell+1) \kappa^{2\ell} \\
 K & = & \frac12 \Bigl[ \frac{1-\kappa^2}{(1-2\kappa\cos(\omega/2)+\kappa^2)2}
   + \frac{1-\kappa^2}{(1+2\kappa\cos(\omega/2)+\kappa^2)2}\Bigr] \\
{\mathcal R} K & = & \frac{1-\kappa^4}{(1-2\kappa^2\cos\eta+\kappa^4)^{3/2}} \\
{\mathcal R} {\mathcal R} K & = & \frac2{\pi} \frac{1-\kappa^2}{(C-D)\sqrt{C+D}} \mbox{E}(\frac{2D}{C+D})
\end{eqnarray*}
where $C = 1-2\kappa\cos\eta_1\cos\eta_2+\kappa^2$ and $D = 2\kappa\sin\eta_1\sin\eta_2$

Analogously, for the de la Vall\'ee Poussin kernel
\begin{eqnarray*}
a_{\ell} & = & (2\mbox{B}(\frac32,\kappa+\frac12))^{-1}[S_{\ell} (\kappa)-S_{\ell+1}(\kappa)]
\end{eqnarray*}
where $S_{\ell}(\kappa) = \sum_{k=0}^{\ell} (-1)^k {2\ell \choose 2k} \mbox{B}(k+\frac12,\kappa+\ell-k+\frac12)$,
and further
\begin{eqnarray*}
K & = & \frac{\mbox{B}(3/2,1/2)}{\mbox{B}(3/2,\kappa+1/2)} \cos(\omega/2)^{2\kappa} \\
{\mathcal R} K & = & (1+\kappa) \cos(\eta/2)^{2\kappa} \\
{\mathcal R} {\mathcal R} K & = & \frac1{\pi} \frac{2^{1-\kappa}}{\cos \eta_1 \cos \eta_2}
 \frac{\Gamma(2+\kappa)}{\Gamma(\frac 32 + \kappa)} \\
& & \qquad \Bigl( A^{1+\kappa} {_2\mbox{F}_1}(\frac12,1+\!\kappa,\frac32+\!\kappa,\frac AB) -
  B^{1+\kappa} {_2\mbox{F}_1}(\frac12,1+\!\kappa,\frac32+\!\kappa,\frac BA) \Bigr)
\end{eqnarray*}
where $A = 1+\cos(\eta_1 + \eta_2)$ and $B = 1+\cos(\eta_1 - \eta_2)$.

\section{Conclusions}
The essential role of the probability density of the angle distribution for the inverse spherical ``totally geodesic'' Radon transform has been clarified by purely geometric arguments. It is identified with the generalized spherical Radon transform which in turn is identified with the ``spherically translated'' spherical Radon transform. Of particular interest is that the twofold spherical Radon transform of a symmetrical kernel function is again its corresponding angle probability density function. Practical methods of inversion in terms of harmonics or radially symmetric basis functions are sketched. Thus, our contribution is also a tribute to the late Hans--Joachim Bunge (1929 -- 2004), who introduced the angle distribution into ``quantitative texure analysis''.
The problem whether the inversion of the generalized spherical Radon transform is better conditioned than the inversion of the spherical radon transform is postponed to a future contribution as it requires a detailed analysis of the experiment to collect integral radiation intensity data.

\section{Acknowledgment}
The authors SB and RH gratefully acknowledge financial support by Deutsche Forschungsgemeinschaft, grant ``high resolution texture analysis'' (SCHA 465/15).
This contribution also relates to the European Union INTAS project ``Investigation of crystallographic textures and elastic and plastic anisotropy for materials with hexagonal symmetry'' (INTAS Ref. Nr. 03-51-6092), the partners of which are (i) University of Metz, France, (ii) All-Russian Institute of Aviation Materials, Moscow, Russia, (iii) Moscow Engineering Physics Institute, Mscow, Russia, (iv) Joint Institute of Nuclear Research, Dubna, Russia, and (v) Freiberg University of Mining and Technology, Germany.

\end{document}